\documentclass[aps,prd,floatfix,superscriptaddress,showpacs,showkeys]{revtex4}
\usepackage{amssymb}
\usepackage{amsmath}
\usepackage{amsfonts}
\usepackage{epsfig}
\usepackage{graphicx}
\usepackage{tabularx}

\usepackage{latexsym}
\usepackage{graphics}
\usepackage{verbatim}
\usepackage{color}

\def\eq{\begin{eqnarray}}
\def\en{\end{eqnarray}}

\def\L{{\cal L}}

\def\Lam{\Lambda_c}

\def\Lam+{\Lambda_c^+}
\def\Lamp+{\Lambda_c^{'+}}
\def\blam+{{\bar{\Lambda}_c^+}}

\def\bp{\bar{p}}
\def\lcp{\Lambda_c(2940)}

\def\lc{\Lambda_c(2286)}
\def\blc{\bar\Lambda_c(2286)}

\def\L2{\Lambda^2}

\begin{document}
\title{Role of the hadron molecule $\Lambda_c(2940)$ \\
in the $p\bar p \to  pD^0\bar\Lambda_c(2286)$ annihilation reaction\\}
\noindent
\author{Yubing Dong}
\affiliation{Institute of High Energy Physics, Beijing 100049, 
People's Republic China} 
\affiliation{Theoretical Physics Center for Science Facilities (TPCSF), CAS,
Beijing 100049, People's Republic China}
\author{Amand  Faessler}
\affiliation{Institut f\"ur Theoretische Physik,  Universit\"at T\"ubingen,\\
Kepler Center for Astro and Particle Physics, \\
Auf der Morgenstelle 14, D--72076 T\"ubingen, Germany}
\author{Thomas Gutsche}
\affiliation{Institut f\"ur Theoretische Physik,  Universit\"at T\"ubingen,\\
Kepler Center for Astro and Particle Physics, \\
Auf der Morgenstelle 14, D--72076 T\"ubingen, Germany}
\author{Valery E. Lyubovitskij}
\affiliation{Institut f\"ur Theoretische Physik,  Universit\"at T\"ubingen,\\
Kepler Center for Astro and Particle Physics, \\
Auf der Morgenstelle 14, D--72076 T\"ubingen, Germany}
\affiliation{Department of Physics, Tomsk State University,
634050 Tomsk, Russia}
\affiliation{Mathematical Physics Department,
Tomsk Polytechnic University, Lenin Avenue 30,
634050 Tomsk, Russia}

\date{\today}

\begin{abstract}

The annihilation process $p\bar p \to pD^0\bar\Lambda_c(2286)$
is studied taking into account $t$-channel $D^0$, $D^{\ast 0}$
meson exchange and the resonance contribution of $\Lambda_c(2286)$ 
and $\Lambda_c(2940)$ baryons. 
We assume that the $\Lambda_c(2940)$ baryon is a $pD^{\ast 0}$
molecular state with spin-parity $\frac{1}{2}^+$ and $\frac{1}{2}^-$.  
Our results show that near the threshold of 
$p\bp\to \Lambda_c(2286)\bar\Lambda_c(2286)$
the contribution from the intermediate state $\Lambda_c(2940)$ is also
sizeable and can be observed at the $\bar{P}$ANDA experiment. 
Another conclusion is that 
the spin-parity assignment $\frac{1}{2}^-$ for $\Lambda_c(2940)$ gives 
enhancement for the cross section in comparison with a 
choice  $\frac{1}{2}^+$. 

\end{abstract}


\maketitle

\section{Introduction}

Studies of the nucleon, nucleon excitations and other baryon resonances with
heavy quarks are of great interest in exploring the structure of hadrons.
Many related experiments with the aim to investigate baryon resonances have
been carried out at facilities like JLab,
BEPC, {\it BABAR} and Belle etc., by using lepton probes as well
as $e^+e^-$ scattering techniques.
The experiments based on the $p\bp$ annihilation process provide
another way to produce heavy baryon resonances which are detected
in various decay channels. Forthcoming experiments at $\bar{P}$ANDA, 
with the $\bar{p}$ momentum in the range from 1 to 15 GeV/c, which
corresponds to total center-of-mass energies in the antiproton-proton 
system between 2.25 and 5.5 GeV, can give rich contributions to 
these investigations~\cite{Elisa}. 
For example, $p\bar p $ annihilation reactions are expected to 
provide substantial information on the charm baryon $\lc$ as well as the 
baryon resonance  $\lcp$ recently
observed by the {\it BABAR} Collaboration~\cite{Aubert:2006sp} and 
confirmed by the Belle Collaboration~\cite{Abe:2006rz}.

Theoretical studies on the $\lcp$ state have been done 
assuming different assignments for its spin-parity  
$J^P = \frac{1}{2}^\pm, \frac{3}{2}^\pm, \frac{5}{2}^\pm$  
and within different approaches~\cite{He:2006is}-\cite{Dong:2009tg} 
(for an overview see Ref.~\cite{Dong:2009tg}). 
In Ref.~\cite{Hejun} it was discussed the
production rate of $\lcp$ at the forthcoming $\bar{P}$ANDA 
experiment based the different assignments for the $\Lambda_c(2940)$
spin-parity. It is a first calculation for the total cross section but
for example initial state interaction and the contribution of $D^\ast$ 
meson exchange are not considered. 

In this work we study the resonance $\lcp$ as a $(pD^\ast)$ hadronic 
molecular state with the help of a phenomenological Lagrangian approach. 
In our previous analysis~\cite{Dong:2009tg} 
of the strong two-body decays of the $\lcp$ we showed that 
its spin-parity assignment $J^P=\frac{1}{2}^+$ is favored. 
This ansatz for the $\lcp$ has been proved to be also reasonable for 
the observed modes in three-body and radiative
decays~\cite{Dong:2009tg}.
Here for completeness we also consider the $J^P=\frac{1}{2}^-$ assignment. 
The technique for describing and treating composite hadron systems was
for example already shown in Refs.~\cite{Dong:2009tg}-\cite{Dong:2008gb}. 
Here we aim for a
quantitative determination of a production mode of the $\lcp$, we
determine cross sections for the annihilation process $p\bp \to
\lcp\to pD^0 \blc$. It is expected that in experiments of
$\bar{P}$ANDA these quantities can possibly be measured. Our
predictions together with the structure assumption can provide additional
information on the nature of this new resonance.

The paper is organized as follows. In Sec.~II, we will briefly
discuss the effective Lagrangian approach for the couplings of
$\lc\to pD^0$ and $\lcp\to pD^{\ast 0}$. Then we introduce the relevant
theory elements to describe the transition $p\bp \to \lcp\blc \to
pD^0 \blc$. Section III is devoted to the numerical results for the
differential and total cross section of $p\bp \to \lcp \blc \to pD^0
\blc$. In the calculation we take initial state interaction  as well
as the $D$ and $D^\ast$ meson exchange $t$--channel contributions
into account. Finally, we briefly summarize our results.

\section{Approach}

We consider two assignments for 
the spin and parity quantum numbers of
the $\lcp$ --- $J^{\rm P} = \frac{1}{2}^{+}$ and 
$J^{\rm P} = \frac{1}{2}^{-}$. 
While the $\frac{1}{2}^{+}$ assignment is favored in our analysis of
the strong decays of the $\lcp$ here we consider both possibilities 
for $J^{\rm P}$. We consider this resonance as a bound state dominated by
the molecular $p D^{\ast 0}$ component
\eq
\label{Lcstate}
|\lcp\rangle  \, = \,  | p D^{\ast 0}\rangle \,.
\en
The annihilation processes $p\bar p \to \lc \blc \to pD^0 \blc$
and $p\bar p \to \lcp \blc \to pD^0 \blc$ are described by 
$t$--channel diagrams
based on the exchange of $D$ and $D^\ast$ mesons (see Fig.~1).
The evaluation of the Feynman diagrams relies on several
elements for the effective interaction of the involved hadrons.
In the following we use the following 
notations in the formulas with $\Lambda_c(2286) \equiv \Lambda_c$ 
and $\Lambda_c(2940) \equiv \Lambda_c'$. 

\begin{figure}[hb]
\centering
\includegraphics [scale=0.8]{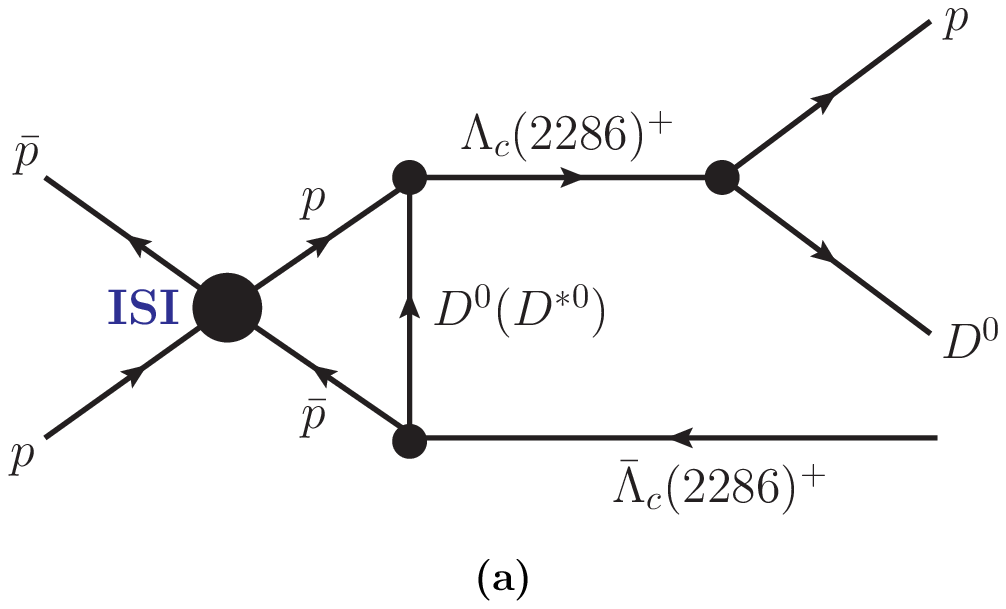}
\hspace{.25cm}
\includegraphics [scale=0.8]{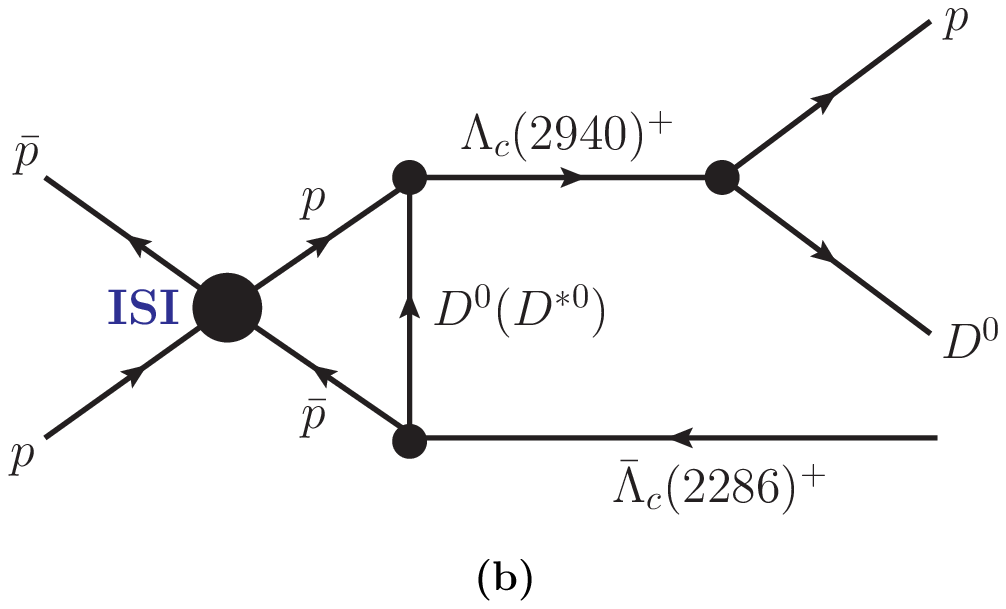}
\caption{$D^0$ and $D^{\ast }$ meson exchange diagrams
contributing to the $P\bar{p}\to pD^0\bar{\Lambda}_c$
process: (a) resonance contribution of the $\lc$ and
         (b) of the $\lcp$ baryon."ISI" in the figures stands for
the initial state interaction.}
\end{figure}

The couplings $g_{BpD}$, $g_{BpD^\ast}$, defining the $BpD$ and
$BpD^\ast$ interactions (where $B = \Lambda_c, \Lambda_c'$), enter in the
phenomenological interaction Lagrangians involving the $\lc$ baryon with 
\eq
{\cal L}_{\Lambda_c pD} &=&
g_{_{\Lambda_c pD}} \, \bar\Lambda_c \, i\gamma_5 \, p \, D^0
\, + \, {\rm H.c.}\,, \label{lbd1}\\
{\cal L}_{_{\Lambda_c pD^\ast}} &=& g_{_{\Lambda_c pD^\ast}} \, 
\bar\Lambda_c \, 
\gamma^{\mu} \,p \, D^{\ast 0}_{\mu} \, + \, {\rm H.c.} \label{lbd2}
\en 
The ones involving the $\Lambda_c'$ baryon for the  
$J^{\rm P} = \frac{1}{2}^{+}$ and 
$J^{\rm P} = \frac{1}{2}^{-}$ assignments are set up as
\eq
{\cal L}_{\Lambda_c' pD}^{\frac{1}{2}^+} &=&
g_{_{\Lambda_c' pD}} \, \bar\Lambda_c' \, i\gamma_5 \, p \, D^0
\, + \, {\rm H.c.}\,, \label{lbdp1}\\
{\cal L}_{_{\Lambda_c' pD^\ast}}^{\frac{1}{2}^+} &=& 
g_{_{\Lambda_c' pD^\ast}} \, \bar\Lambda_c' \, 
\gamma^{\mu} \,p \, D^{\ast 0}_{\mu} \, + \, {\rm H.c.} \label{lbdp2}
\en 
and 
\eq
{\cal L}_{\Lambda_c' pD}^{\frac{1}{2}^-} &=&
f_{_{\Lambda_c' pD}} \, \bar\Lambda_c' \, p \, D^0
\, + \, {\rm H.c.}\,, \label{lbfp1}\\
{\cal L}_{_{\Lambda_c' pD^\ast}}^{\frac{1}{2}^-} &=& 
f_{_{\Lambda_c' pD^\ast}} \, \bar\Lambda_c' \, 
\gamma^{\mu} \gamma^5\,p \, D^{\ast 0}_{\mu} 
\, + \, {\rm H.c.} \label{lbfp2}\,. 
\en 
The couplings in these Lagrangians 
have been determined in Ref.~\cite{Dong:2009tg}. 
In particular, from $SU(4)$
invariant Lagrangians~\cite{Dong:2009tg,SU4} we deduce the couplings
\eq g_{_{\Lambda_c pD}}  = - \frac{3\sqrt{3}}{5} \, g_{_{\pi NN}} = - 14.97
\,, \quad g_{_{\Lambda_c pD^\ast}} = - \frac{\sqrt{3}}{2} \, 
g_{_{\rho NN}} = - 5.20  \, \en given in terms of the pion-nucleon 
$g_{_{\pi NN}} = 13.4$ and the vector rho-nucleon 
$g_{_{\rho NN}} = 6$ coupling constants.

The couplings $g_{\Lambda_c' pD}$, $g_{\Lambda_c' pD^\ast}$ and 
$f_{\Lambda_c' pD}$, $f_{\Lambda_c' pD^\ast}$ have been
evaluated in the hadronic molecular
approach~\cite{Faessler:2007gv,Dong:2009tg} for the $\lcp$ baryon
state using the compositeness
condition~\cite{Weinberg:1962hj}-\cite{Anikin:1995cf} with  
\eq
& &g_{_{\Lambda_c' pD}} = - 0.54 \,, \quad\quad 
g_{_{\Lambda_c' pD^\ast}} = 6.64\,, 
\nonumber\\
& &f_{_{\Lambda_c' pD}} = - 0.97 \,, \quad\quad 
f_{_{\Lambda_c' pD^\ast}} = 3.75 \,. 
\en
The dressed $M = D, D^\ast$ meson propagators are accompanied by the
vertex form factors \eq
F_{_{M}}(t)=\frac{\Lambda^2_{_{M}}-M^2_{_{M}}} {\Lambda^2_{_{M}}+t}
\en encoding the off shellness of $D(D^\ast)$ mesons, where
$\Lambda_{_{M}} =  3$ GeV is the cutoff parameter and $t$ stands for
the exchanged momentum squared~\cite{Holzenkamp:1989tq}. 
When choosing the cutoff parameter as $\Lambda_{_{M}} =  3$ GeV we follow 
the argument given in Ref.~\cite{Haidenbauer:2009ad}, where such a value was 
originally used. 
As was found~\cite{Haidenbauer:2009ad}  
the cross section for $p\bar p \to  \Lambda_c(2286)\bar\Lambda_c(2286)$  
is sensitive to a variation of the parameter $\Lambda_{_{M}}$ and 
reduces by a factor 3 when $\Lambda_{_{M}}$ decreases from 3 to 2.5 GeV. 
It was pointed out~\cite{Haidenbauer:2009ad} that the value of the cutoff 
parameter $\Lambda_{_{M}}$ should be bigger than either one of
the masses of the exchanged charmed mesons $D$ and $D^\ast$. 

Note that we performed a microscopic calculation for $g_{\Lambda_c' pD}$, 
$g_{\Lambda_c' pD^\ast}$ and $f_{\Lambda_c' pD}$, 
$f_{\Lambda_c' pD^\ast}$ based on the molecular structure 
of the $\lcp$ state with a clear dominance (by a factor $\simeq 3$) of 
the $f$-couplings corresponding to the $\frac{1}{2}^-$ assignment of 
the $\lcp$ state. Note, in Ref.~\cite{Hejun} such couplings were fixed 
from the two-body decay widths of the $\lcp$ assuming that this widths 
are the same for all spin-parity assignments. Obviously, this procedure 
is not quite consistent because of the different spin-parity structures 
and phase spaces. 

The intermediate $\lc$ and $\lcp$ baryon resonances
are described by a Breit--Wigner form contained in the propagators
with a constant width $\Gamma_B$ in the imaginary part:
\begin{equation}
S_B(p) = \frac{M_B+\not\! p}{M^2_B-p^2-i M_B\Gamma_B} \,,
\quad\quad B = \Lambda_c, \Lambda_c'  \,,
\end{equation} 
where $\Gamma_{\Lambda_c} \simeq 3.3 \times 10^{-9}$ MeV 
and $\Gamma_{\Lambda_c'} = 17^{+8}_{-6}$ MeV 
are the  widths of the $\Lambda_c(2286)$ and $\Lambda_c(2940)$ states, 
respectively. In our calculation we use the central value of 
$17$ MeV for the width of $\Lambda_c(2940)$.   

Following Ref.~\cite{Haidenbauer:2009ad}
we also take into account the initial state interaction (ISI) for the $p\bp$
entrance channel.
For the $T$ matrix of the $N\bar{N}$ interaction we use the
Lippmann-Schwinger equation
\eq\label{Tpps}
T(\vec{q}^{~'},\vec{q\,};~E)
=V(\vec{q}^{~'},\vec{q\,};~E)+\int \frac{d^3p \, V(\vec{q}^{~'},\vec{p\,})
\, T(\vec{p},\vec{q\,};~E)}{E(q)-E(p)+i\epsilon} \;,
\en
as illustrated in Fig.~2. In above equation
$V_{N\bar{N}}(\vec{q}^{\,'},\vec{q\,})$ is
a phenomenological nucleon-antinucleon potential given by the
sum of a pion exchange $V_{N\bar{N}}^{\pi}(\vec{q}^{\,'},\vec{q\,})$ and
optical nucleon-antinucleon potential
$V_{N\bar{N}}^{\rm opt}(\vec{q}^{\,'},\vec{q\,})$
\eq
V_{N\bar{N}}(\vec{q}^{\,\,'},\vec{q\,}) =
V_{N\bar{N}}^{\pi}(\vec{q}^{\,\,'},\vec{q}\,)
+V_{N\bar{N}}^{\rm opt}(\vec{q}^{\,\,'},\vec{q}\,) \,.
\en
\begin{figure}[hb]
\centering
\includegraphics [scale=0.7]{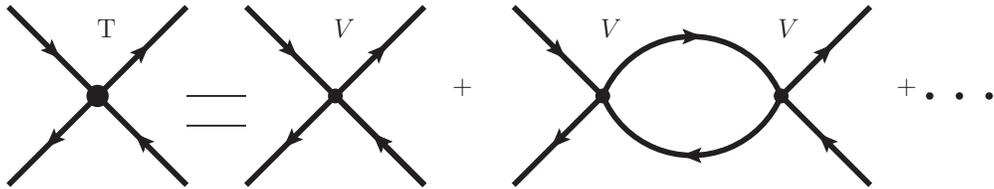}
\caption{Lippmann-Schwinger equation for the initial
state interaction of the $N\bar{N}$ system}
\end{figure}

The $\pi$-exchange potential is given by~\cite{Holzenkamp:1989tq,
Haidenbauer:2009ad}
\eq
V_{N\bar{N}}^{\pi}(\vec{q}^{\,'},\vec{q}\,) =
\frac{g_{\pi NN}^2}{12 M_N^2} \,
\frac{\vec{k}^{\,2}_{\pi}}{M_\pi^2 + \vec{k}^{\,2}_{\pi}} \,
\Big( \vec{\sigma}_1 \cdot \vec{\sigma}_2 \, + \,
\hat{S}_{12}(\vec{k}_{\pi})\Big) \ (\vec{\tau}_1 \cdot \vec{\tau}_2) \
F_\pi^2(\vec{k}_{\pi}^{\,2}) \,,
\en
with $M_N$ and $M_{\pi}$ being the masses of nucleon and pion,
respectively. The potential contains the tensor operator
\eq
\hat{S}_{12}(\vec{k}_{\pi})=3\vec{\sigma}_N\cdot\hat{\vec{k}}_{\pi}
\vec{\sigma}_{\bar{N}}\cdot\hat{\vec{k}}_{\pi}-
\vec{\sigma}_N\cdot\vec{\sigma}_{\bar{N}},
\en
where $\hat{\vec{k}}=\vec{k}/\mid\vec{k}\mid$ and
$\vec{k}_{\pi}=\vec{q}-\vec{q}^{~'}$ is the three-momentum
of the pion. $F_\pi(\vec{k}_{\pi}^{\,2})$ is a phenomenological monopole
form factor
\eq
F_{\pi}(\vec{k}_{\pi}^{\,2})=\frac{\Lambda^2_{\pi}-M^2_{\pi}}
{\Lambda_{\pi}^2+\vec{k}_\pi^2},
\en
where $\Lambda_{\pi} = 1.3$ GeV is the cutoff parameter.

The optical potential for the $N\bar N$ scattering state is given
by~\cite{Holzenkamp:1989tq,Haidenbauer:2009ad}
\eq
V_{N\bar{N}}^{\rm opt}(r) \, = \, (u_0+iw_0) \, e^{-\vec{r}^{~2}/2r_0^2}
\en
where the parameters were fixed as $u_0 = -0.0480$ GeV,
$w_0 = 0.5319$ GeV and $r_0=0.56$ fm.

For the calculation of the process in Fig.~1 we assume that the
ISI can be factorized out by the dimensionless factor
\eq
\label{ISI}
J_0 &=&\int d^3q' \, T_{N\bar{N}}(\vec{q}^{\,'},\vec{q}) \,
\frac{1}{E_{p_1}+E_{P-p_1}-\sqrt{s}+i\epsilon} \,.
\en

\vspace*{1cm }
\begin{figure}[htb]
\centering
\includegraphics[scale=0.35]{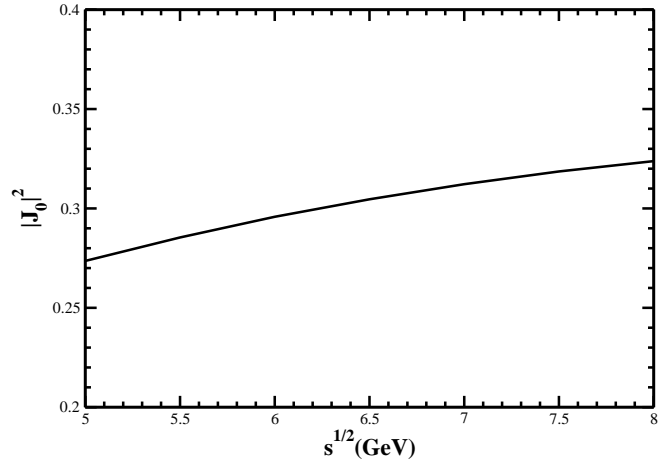}

\caption{Initial state interaction factor $\mid J_0\mid^2$
in dependence on $s^{1/2}$}
\end{figure}

In the evaluation of $J_0$ we use the center-of-momentum frame, 
where the momenta of
incoming ($p_1,p_2=P-p_1$) and outgoing ($p_1',P-p_1'$) particles
are defined as 
\eq p_1  = (E_1,  \vec{q\,})\,, \quad p_2  =
(E_2, -\vec{q\,})\,, \quad p_1' = (E_1', \vec{q}^{\,'})\,,
\quad p_2' = (E_2',-\vec{q}^{\,'})\,,
\en
and $s = (E_1 + E_2)^2$ is the total energy squared.

Finally, the invariant matrix element corresponding to the process 
$p \bar p \to p D^0 \bar\Lambda(2286)$ is written as 
\eq 
{\cal M}_{\rm inv} &=& 
{\cal M}_{\rm inv}^{(a)} + {\cal M}_{\rm inv}^{(b)} \,. 
\en 
${\cal M}_{\rm inv}^{(a)}$ is the contribution of 
diagram in Fig.1(a) [contribution of the $\Lambda_c(2286)$ state] 
\eq 
{\cal M}_{\rm inv}^{(a)} &=& 
g_{\rm eff}^{aP} \frac{F_M^2(t)}{M_D^2-t} \, \bar{u}(q_1) i\gamma_5
\frac{M_{\Lambda_c}+\not\! p_4}{M_{\Lambda_c}^2-p_4^2 
-iM_{\Lambda_c}\Gamma_{\Lambda_c}}i\gamma_5 u(p_1) 
\ \bar{v}(p_2)i\gamma_5v(q_2)
\nonumber\\
&+&g_{\rm eff}^{aV} \frac{F_M^2(t)}{M^2_{D^{\ast}}-t} \, 
\biggl( - g^{\mu\nu} + \frac{p_3^\mu p_3^\nu}{M^2_{D^{\ast}}}\biggr) 
\bar{u}(q_1) i\gamma_5
\frac{M_{\Lambda_c}+\not\! p_4}{M_{\Lambda_c}^2-p_4^2 
-iM_{\Lambda_c}\Gamma_{\Lambda_c}} \gamma_\mu u(p_1) 
\ \bar{v}(p_2)\gamma_\nu v(q_2) \,.
\en 
The amplitude ${\cal M}_{\rm inv}^{(b)}$ is the result of the
diagram in Fig.~1(b) [contribution of the $\Lambda_c(2940)$ state]. 
For assignments $\frac{1}{2}^+$ and $\frac{1}{2}^-$ it is given by  
\eq 
& &\mathrm{assignment} \ J^P = \frac{1}{2}^+\nonumber\\ 
{\cal M}_{\rm inv}^{(b)} &=& 
g_{\rm eff}^{bP} \frac{F_M^2(t)}{M_D^2-t} \, \bar{u}(q_1) i\gamma_5
\frac{M_{\Lambda_c'}+\not\! p_4}{M_{\Lambda_c'}^2-p_4^2 
-iM_{\Lambda_c'}\Gamma_{\Lambda_c'}}i\gamma_5 u(p_1) 
\ \bar{v}(p_2)i\gamma_5v(q_2)
\nonumber\\
&+&g_{\rm eff}^{bV} \frac{F_M^2(t)}{M^2_{D^{\ast}}-t} \, 
\biggl( - g^{\mu\nu} + \frac{p_3^\mu p_3^\nu}{M^2_{D^{\ast}}}\biggr) 
\bar{u}(q_1) i\gamma_5
\frac{M_{\Lambda_c'}+\not\! p_4}{M_{\Lambda_c'}^2-p_4^2 
-iM_{\Lambda_c'}\Gamma_{\Lambda_c'}} \gamma_\mu u(p_1) 
\ \bar{v}(p_2)\gamma_\nu v(q_2)
\en 
and 
\eq 
& &\mathrm{Assignment} \ J^P = \frac{1}{2}^-\nonumber\\ 
{\cal M}_{\rm inv}^{(b)} &=& 
f_{\rm eff}^{bP} \frac{F_M^2(t)}{M_D^2-t} \, \bar{u}(q_1) \, 
\frac{M_{\Lambda_c'}+\not\! p_4}{M_{\Lambda_c'}^2-p_4^2 
-iM_{\Lambda_c'}\Gamma_{\Lambda_c'}}u(p_1) 
\ \bar{v}(p_2) i\gamma_5 v(q_2)
\nonumber\\
&+&f_{\rm eff}^{bV} \frac{F_M^2(t)}{M^2_{D^{\ast}}-t} \, 
\biggl( - g^{\mu\nu} + \frac{p_3^\mu p_3^\nu}{M^2_{D^{\ast}}}\biggr) 
\bar{u}(q_1) i\gamma_5
\frac{M_{\Lambda_c'}+\not\! p_4}{M_{\Lambda_c'}^2-p_4^2 
-iM_{\Lambda_c'}\Gamma_{\Lambda_c'}} \gamma_\mu \gamma_5 u(p_1) 
\ \bar{v}(p_2)\gamma_\nu v(q_2)\,.
\en 
Here we use the following notations: 
$p_1$, $p_2$, $p_3$, $q_1$, $q_2$, $q_3$ are the momenta of 
initial proton, initial antiproton, the exchanged $D^0(D^{0\ast})$ meson, 
final proton, final $\bar\Lambda(2286)$ and 
final $D^0$ meson, respectively; $t = p_3^2$;  
$p_4 = q_1 + q_3 = M_{pD}$ is the momentum of the $\lc$ $(\lcp)$ resonance 
related to the invariant mass of the final proton and $D^0$ meson;  
$u(p_1)$, $\bar v(p_2)$, $\bar u(q_1)$, $v(q_2)$ are  
the spinors describing initial proton, initial antiproton, final proton and 
final $\bar\Lambda(2286)$, respectively.  
The couplings $g_{\rm eff}^{ij}$ and $f_{\rm eff}^{ij}$ $(i=a,b; j=P,V)$ 
are defined as 
\eq 
& &g_{\rm eff}^{aP} = J_0 \, g_{_{\Lambda_c pD}}^3\,, \quad 
   g_{\rm eff}^{aV} = J_0 \, g_{_{\Lambda_c pD}} \, 
   g_{_{\Lambda_c pD^\ast}}^2\,, 
\nonumber\\
& &g_{\rm eff}^{bP} = J_0 \, g_{_{\Lambda_c' pD}}^3\,, \quad 
   g_{\rm eff}^{bV} = J_0 \, g_{_{\Lambda_c' pD}} \, 
   g_{_{\Lambda_c' pD^\ast}}^2\,,\\
& &f_{\rm eff}^{bP} = J_0 \, g_{_{\Lambda_c' pD}} \, 
   f_{_{\Lambda_c' pD}}^2\,, \quad 
   f_{\rm eff}^{bV} = J_0 \, g_{_{\Lambda_c' pD}} \, 
   f_{_{\Lambda_c' pD^\ast}}^2\,. 
\en

\section{Numerical results and discussion}

The differential cross section for the process $p\bp\to
pD^0\bar{\Lambda}_c$ is obtained through the expression 
\eq
\frac{d\sigma}{dM_{pD}} = 
\frac{1}{1024 \pi^4}\frac{1}{s \, \sqrt{s-4M_N^2}} \, 
\int d\!\cos\theta_3 \ d\Omega^\ast_1 \ 
\mid\vec{q}_1^{~\ast}\mid 
\ \mid\vec{q}_2\mid \ {\mid{\cal M}_{\rm inv}\mid}^2 
\en 
where $\vec{q}_1^{\,\ast}$ and
$\Omega_1^\ast$ are the three-momentum and solid angle of the outgoing
proton in the center-of-mass frame of the final $pD$ system;
$\vec{q}_2$ and $\theta_2$ are the three-momentum and scattering
angle of the final $\bar{\Lambda}_c(2286)$ state. In
above equation $M_{pD}$ is the invariant mass of the final $pD$
two-body system. The transition amplitude for $p\bp\to
pD^0\bar{\Lambda}_c$ of Fig.~1 is contained in the invariant matrix
element ${\cal M}_{\rm inv}$. The contributions of $D$ and $D^\ast$
exchange as well as of the possible intermediate states $\lc$ and
$\lcp$ are fully taken into account. Masses of the intermediate
baryons and of the exchanged $D$ mesons are taken from the Particle 
Data Group compilation. The effect of initial state interaction is expressed
through the factor $J_0$ of Eq.~(\ref{ISI}) os also present in 
${\cal M}_{\rm inv}$. Neglecting ISI would correspond to $J_0=1$. Values for
$\mid J_0\mid^2$ are displayed in Fig.~3 indicating a sizable
suppression of the transition as induced by ISI.

\begin{figure}
\includegraphics   [scale=0.41]{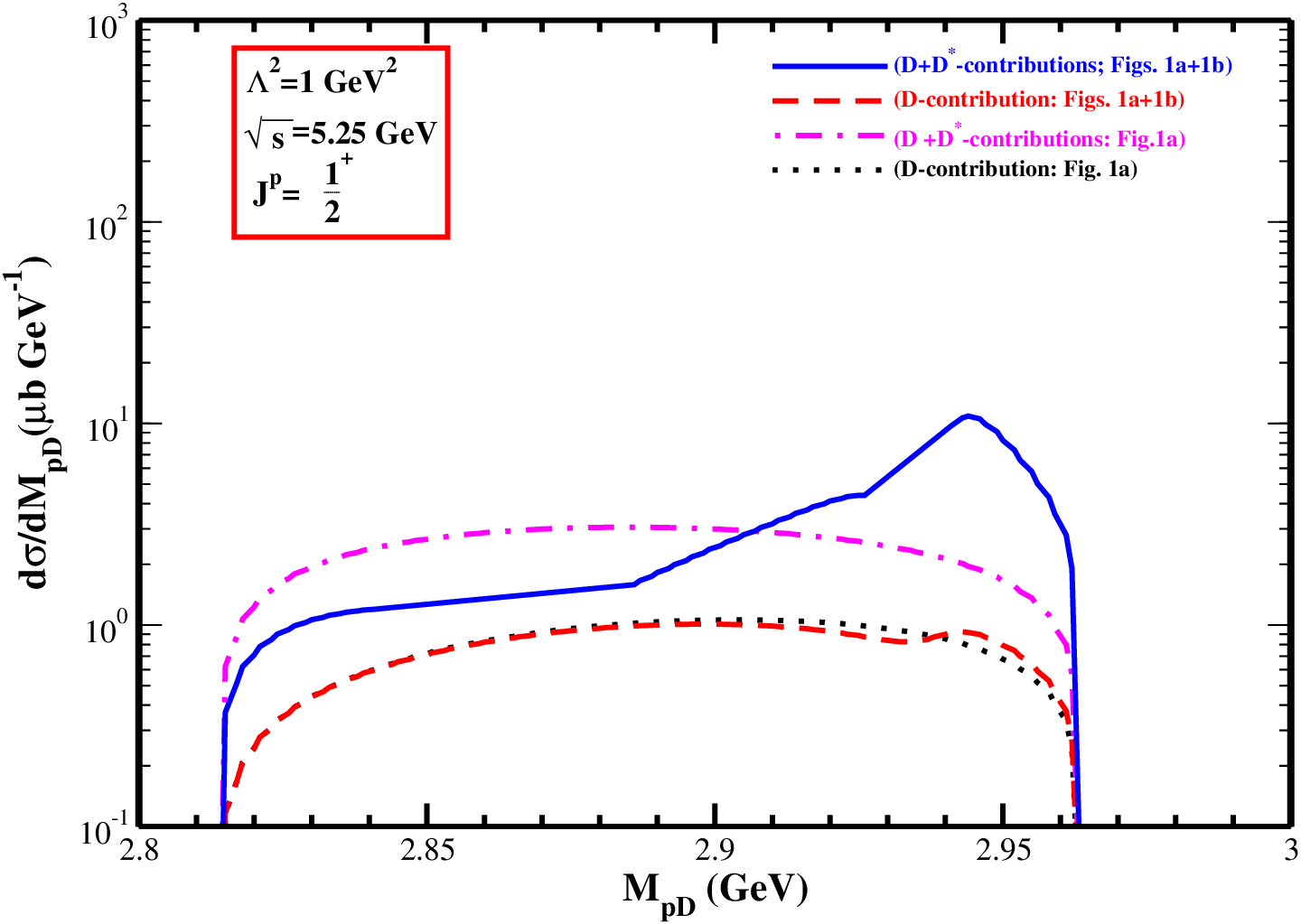}
\caption{Differential cross section $d\sigma/dM_{pD}$
for $s^{1/2}=5.25$~GeV for $J^P = \frac{1}{2}^+$ of 
the $\Lambda_c(2940)$}

\vspace*{1.2cm}

\includegraphics   [scale=0.41]{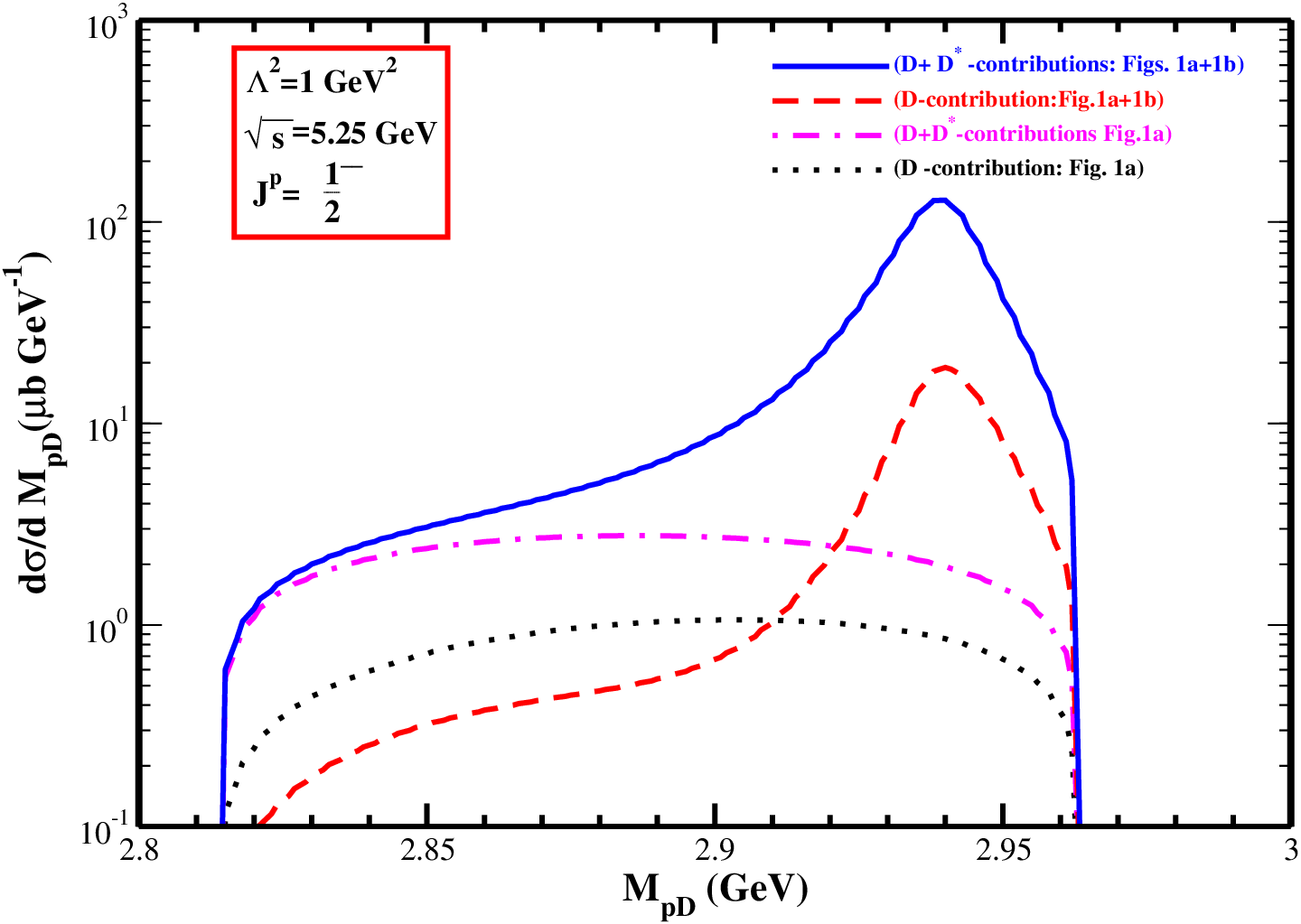}
\caption{Differential cross section $d\sigma/dM_{pD}$
for $s^{1/2}=5.25$~GeV for $J^P = \frac{1}{2}^-$ of 
the $\Lambda_c(2940)$}
\end{figure}

\begin{figure}
\includegraphics   [scale=0.41]{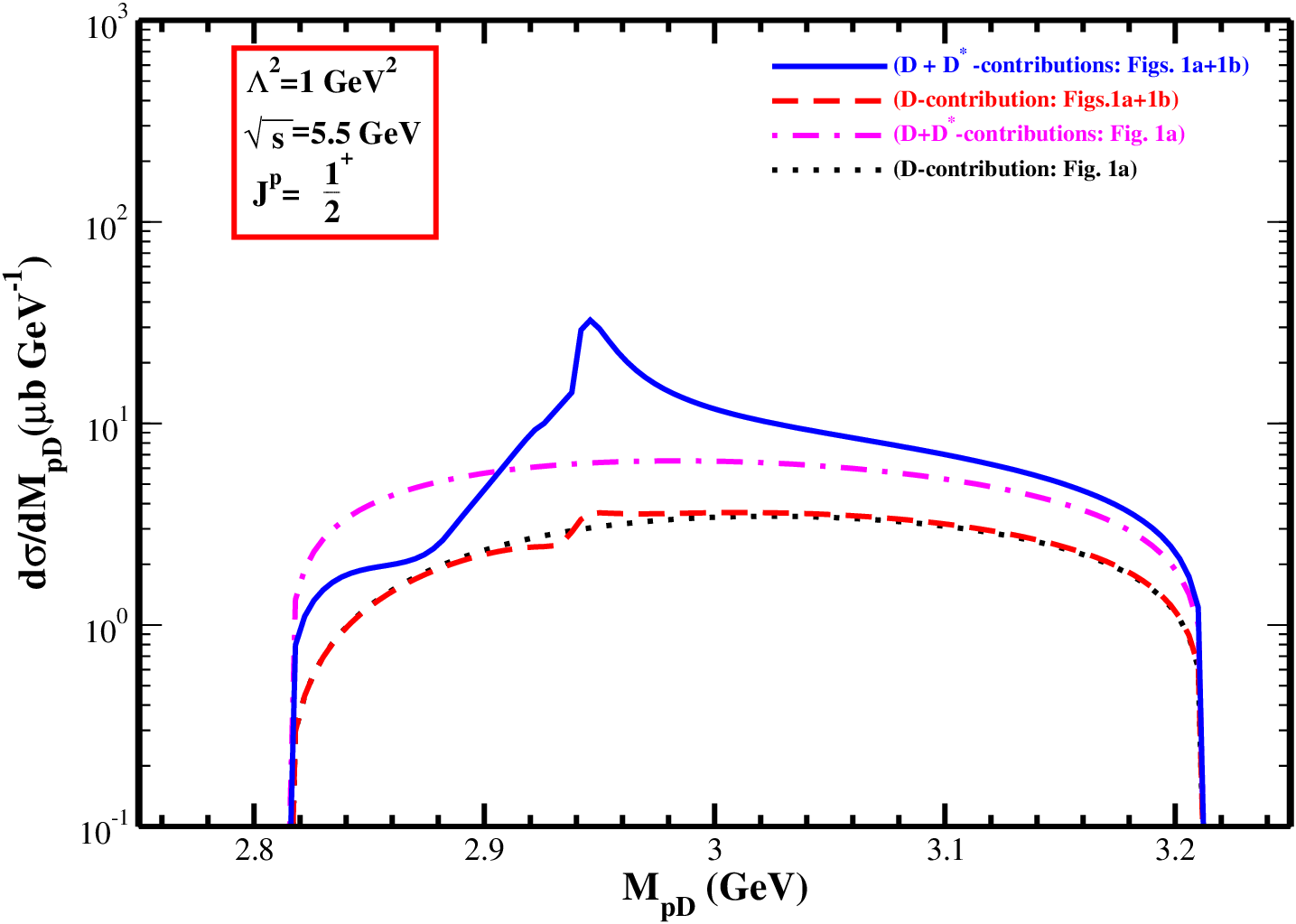}
\caption{Differential cross section $d\sigma/dM_{pD}$
for $s^{1/2}=5.5$~GeV for $J^P = \frac{1}{2}^+$ of 
the $\Lambda_c(2940)$}

\vspace*{1.2cm}

\includegraphics   [scale=0.41]{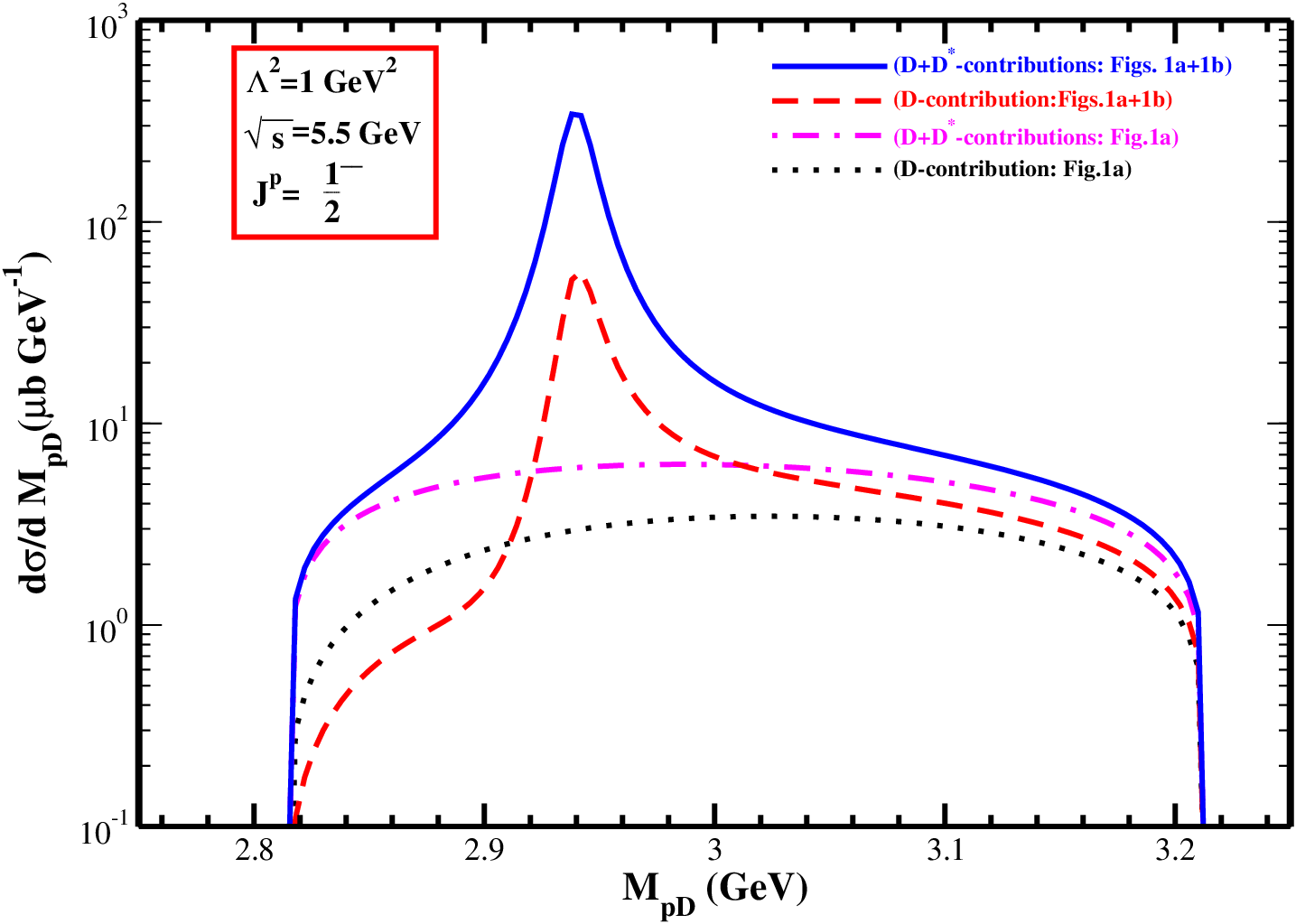}
\caption{Differential cross section $d\sigma/dM_{pD}$
for $s^{1/2}=5.5$~GeV for $J^P = \frac{1}{2}^-$ of 
the $\Lambda_c(2940)$}
\end{figure}

In Figs.~4-7 we show the differential cross sections
$d\sigma/dM_{pD}$ for the total energies $\sqrt{s}=5.25$~GeV 
and $\sqrt{s}=5.5$~GeV. In the calculation we take the $\Lambda_c(2940)$ 
as a hadronic molecule as of Eq.~(\ref{Lcstate}). The size parameter of the 
correlation function is selected as $\Lambda^2=1$ GeV$^2$~\cite{Dong:2009tg}
in the hadron molecule scenario.  
We explicitly display the contributions ---  of the diagram in Fig.~1(a) 
with $D^0$ exchange only (dotted line), of the diagram in Fig.~1(a) with 
$D^0$ and $D^{\ast 0}$ exchange (dot-dashed line), of the 
diagrams in Figs.~1(a) and~1(b) with $D^0$ exchange only (dashed line) 
and the full contribution of the diagrams in Figs.~1(a) and~(1b), including  
both $D^0$ and $D^{\ast 0}$ exchange (solid line).  
A change of the spin-parity assignment from  $\frac{1}{2}^+$ to $\frac{1}{2}^-$ 
leads to an enhancement of the cross section by a factor 10. Also, the $\Lambda_c(2940)$ 
resonance gives a sizable contribution, which can be checked at 
the $\bar{P}$ANDA experiment. 

To summarize, we have estimated the differential and total cross
sections of $p\bp\to pD^0\bar{\Lambda}_c$ in an energy range relevant
for $\bar{P}$ANDA. In our calculations we include initial state
interaction as well as the $D$ and $D^\ast$ exchange dynamics.
The inclusion of ISI leads to a suppression, $D$ exchange dominates the
transition dynamics.
We include the resonance $\lcp$ which is treated as a 
$\frac{1}{2}^+$ or as a $\frac{1}{2}^-$ molecular $pD^{\ast 0}$ state. 
In our analysis we work out and discuss the role of the $\lcp$ 
in comparison to the background effect including the $\lc$.
We hope and expect that future experiments at $\bar{P}$ANDA will
provide a test to our model calculations especially because 
the two spin-parity assignments can be clearly distinguished. 

\begin{acknowledgments}

This work is supported by the DFG under Contract No. LY 114/2-1,  
by Tomsk State University Competitiveness Improvement Program, 
by National Sciences Foundations of China Grants 
No. 11475192, No. 11035006, and No. 11261130, as well as supported, in part,
by the DFG and the NSFC through funds provided to the Sino-German CRC
110 ``Symmetries and the Emergence of Structure in QCD''.
Y.B.D. thanks the Institute of Theoretical
Physics, University of T\"ubingen for the warm hospitality and
thanks for the support from the Alexander von Humboldt Foundation.
Y.B.D. also thanks for a fruitful discussion with Johann Haidenbauer.

\end{acknowledgments}

\end{document}